\documentstyle[preprint,epsfig,aps]{revtex}

\def\be{\begin{equation}}
\def\ee{\end{equation}}
\def\bea{\begin{eqnarray}}
\def\eea{\end{eqnarray}}

\begin{document}

\title{Additional Nucleon Current Contributions to Neutrinoless 
       Double Beta Decay}

\author{ 
F. \v Simkovic$^1$\thanks{{\it Permanent address:} 
Department of Nuclear Physics,  
Comenius University, Mlynsk\'a dolina F1, 
SK--842 15 Bratislava, Slovakia; 
Electronic address: simkovic@fmph.uniba.sk},
G. Pantis$^2$\thanks{Electronic address: gpantis@cc.uoi.gr},
J.D. Vergados$^1$\thanks{{\it Permanent address:} 
Theoretical Physics Section, University of Ioannina,
GR--451 10, Ioannina, Greece;
Electronic address: vergados@cc.uoi.gr},
and Amand Faessler$^1$\thanks{Electronic address: 
amand.faessler@uni-tuebingen.de},
}

\address{
1.  Institute of Theoretical Physics, University of Tuebingen,
D--72076 Tuebingen, Germany\\
2.  Theoretical Physics Section, University of Ioannina,
GR--451 10, Ioannina, Greece 
}

\date{\today}
\maketitle

\begin{abstract}
We have examined the  importance of momentum dependent induced nucleon
currents such as weak-magnetism and 
pseudoscalar couplings to the amplitude of neutrinoless double
beta decay ($0\nu\beta\beta$-decay) in the mechanisms of light and heavy 
Majorana neutrino as well as in that of Majoron emission.  Such effects are
expected to occur in all nuclear  models in the direction of reducing
the light neutrino matrix
elements by about $30\%$. To test this we have performed a calculation of 
the nuclear matrix elements of the experimentally interesting nuclei A = 76, 
82, 96, 100, 116, 128, 130, 136 and 150 within the  
proton-neutron renormalized Quasiparticle
Random Phase Approximation (pn-RQRPA). 
We have found that indeed  such corrections
vary somewhat from nucleus to nucleus, but in all cases they are
greater than $25\%$. In the case of heavy neutrino the effect is much larger 
(a factor of 3).
Combining out results with the best presently available experimental limits on 
the half-life of the $0\nu\beta\beta$-decay  
we have extracted new limits on the
effective neutrino mass (light and heavy) and the effective Majoron coupling
constant.
\\\\
{PACS numbers:23.40.Hc,21.60.Jz,27.50.+e,27.60.+j}\\\\
\end{abstract}
\section{Introduction}

The neutrinoless double beta decay ($0\nu\beta\beta$-decay)
is expected to occur if lepton number conservation is not an exact symmetry
of nature.  It is thus forbidden in the Standard Model (SM) of 
electroweak interaction. The recent Kamiokande results have given 
evidence that the neutrinos are massive particles and one has to go
beyond the SM. To further understand
neutrinos, we must know whether they are Dirac or Majorana particles,
an issue which only double beta decay can decide.
The $0\nu\beta\beta$-decay can be
detectable only if the ordinary beta decay is 
forbidden or suppressed and the neutrino is a
Majorana particle (i.e. identical to its own antiparticle) with non-zero 
mass \cite{HS94,DTK85,Verg86,Tom91}.
The study of the $0\nu\beta\beta$-decay is stimulated by the 
development of grand unified theories (GUT's) and supersymmetric
models (SUSY) 
representing extensions of the $SU(2)_L \otimes U(1)$ SM. 
The GUT's and SUSY offer a variety of mechanisms
which allow the $0\nu\beta\beta$-decay to occur \cite{Moh98}. The best known 
possibility is via the exchange of a Majorana neutrino between the two
decaying neutrons \cite{HS94,DTK85,Verg86,Tom91,PSV96}, but increased
attention is paid to more exotic processes, like the supersymmetric
R-parity violating mechanisms of $0\nu\beta\beta$-decay 
\cite{Moh86,Ver87,WKS97,FKSS97,FKS98a,FKS98b}. 
Recent review articles \cite{FS98,SC98} give  a detailed account of the
latest developments in this field. 

In this contribution we shall discuss the role of induced currents such as
weak-magnetism and pseudoscalar coupling in the calculation of the
$0\nu\beta\beta$-decay amplitude, which enters the neutrino mass as well as
the Majoron emission mechanisms. So far only the axial-vector 
and the vector parts have been considered systematically and in great 
detail \cite{HS94,DTK85,Verg86,Tom91,PSV96,FS98,SC98}. 

The weak-magnetism
and nucleon recoil terms have been considered in the extraction of the neutrino
mass independent parameters associated with right-handed 
current mechanisms of $0\nu\beta\beta$-decay
and were found to be very important in the case of
the $\eta$ parameter \cite{PSV96,tom85}. This is understood since 
the leading contribution in this mechanism is proportional to the lepton 
momenta. 
In the two-neutrino double beta decay ($2\nu\beta\beta$-decay) 
the weak-magnetism term \cite{BARB99}
resulted in a  renormalization of the Gamow-Teller matrix element 
independent of
the nuclear model and lead to a reduction of only 10 percent to the half-life
of  medium heavy nuclei.
In the case of the light neutrino mass mechanism of $0\nu\beta\beta$-decay 
there has been one attempt 
to include such effects resulting \cite{SUHO91} from the recoil term and their 
contribution has been found very small. The weak-magnetism and 
induced pseudoscalar terms have been considered in connection with the 
heavy neutrino mass exchange mechanism in Ref. \cite{SEIL92} and
their importance have been manifested for
$0\nu\beta\beta$-decay of $^{48}Ca$.

To our knowledge the induced pseudoscalar term, which is
equivalent to a modification of the axial current due to PCAC, 
has been ignored in all calculations studying the light Majorana 
neutrino mass mechanism even though
it provides a contribution, which is in fact greater
than, the included in all calculations, vector contribution.  
This has perhaps happened because in the charged-current weak processes the 
current-current interaction, under the assumption of zero neutrino mass, leads
to terms which except the  vector and axial-vector parts \cite{LLE72} are
proportional to the lepton mass squared hence, i.e. they are small.

The induced pseudoscalar term, however, is a real function of the Lorenz scalar
$q^2$ therefore there is reason to expect it to be important. In fact we find
that such corrections are of order $(\vec{q})^2/((\vec{q})^2+m^2_{\pi})$, i.e.
they are important if the average momentum $<q>$ of the exchanged neutrino can
not be neglected in front of the pion mass. In the case of a light intermediate
neutrino the  mean nucleon-nucleon separation is about 2 fm which implies that 
the average momentum $<q>$ is about 100 MeV. This leads
to corrections of about $30\%$. In the case of a heavy neutrino
exchange the mean internucleon distance is considerably smaller 
and the average momentum $<q>$ is supposed to be considerably larger.

We should mention that in the R-parity violating SUSY mechanism  of 
$0\nu\beta\beta$-decay \cite{FKS98a} one has scalar, pseudoscalar and tensor
couplings at the quark level, which, of course, induce analogous couplings
at the nucleon level. 

 The correct nucleon current is important in any calculation of the nuclear
matrix elements, which must be computed precisely in order 
to obtain quantitative answers for the lepton number 
violating parameters from  the results of $0\nu\beta\beta$-decay
experiments. 

 The goal of the present paper is to obtain reliable nuclear matrix
elements by including 
the above refinements in the nucleon current in conjunction with the
recent improvements of QRPA (renormalization effects due to Pauli 
principle corrections \cite{TS95,SSF96}). In particular to see what effects,
if any, the weak-magnetism and pseudoscalar coupling terms will
have on the neutrino mass mechanism as well as on the Majoron emission 
mechanism of $0\nu\beta\beta$-decay. To this end we have performed calculations,
which cover most of the nuclear targets of experimental
interest ($^{76}{{Ge}}$, $^{82}{{Se}}$, $^{96}{{Zr}}$, $^{100}{{Mo}}$,
$^{116}{{Cd}}$, $^{128}{{Te}}$, $^{130}{{Te}}$, $^{136}{{Xe}}$,
$^{150}{{Nd}}$).

The paper is organized as follows: In Sect. II, the basic elements
of the theory of the $0\nu\beta\beta$-decay relevant
to this work are presented. 
In the Sect. III the contributions coming from these induced currents
to the $0\nu\beta\beta$-decay amplitude
are analyzed. Section IV summarizes the basic ingredients of
the proton-neutron RQRPA (pn-RQRPA) method, which will be used 
for nuclear structure studies of the $0\nu\beta\beta$-decay transitions.
In Sect. V we discuss the calculation of nuclear matrix elements
and deduce limits on lepton-number violating parameters. 
Finally, in Sect. VI  we summarize the results and draw some
conclusions.

\section{Theory}

\subsection{Majorana neutrino mass mechanism}

We shall consider the $0\nu\beta\beta$-decay process assuming 
that the  effective beta decay Hamiltonian acquires the form:
\begin{equation}
{\cal H}^\beta = \frac{G_{{F}}}{\sqrt{2}} 
\left[\bar{e} \gamma_\mu (1-\gamma_5) \nu_{{e L}} \right]
J^{\mu \dagger}_L + {h.c.},
\label{eq:1}   
\end{equation}
where $e $ and $\nu_{{e L}}$ are field operators representing
electron and left handed electron neutrino, respectively. 
We suppose that  neutrino mixing does take place according to
\begin{equation}
\nu_{e L}=\sum_{k=light} ~U^L_{ek}~\chi_{kL} +
\sum_{k=heavy} ~U^L_{ek}~N_{kL},
\label{eq:2}   
\end{equation}
where, $\chi_{k}$ ($N_{k}$)
are fields of light (heavy) Majorana neutrinos with masses
$m_k$ ($m_k << 1$ MeV) and $M_k$ ($M_k >> 1$ GeV), respectively, 
and $U^L_{ek}$ is a unitary mixing matrix. In the first and second terms 
on the r.h.s. of Eq. (\ref{eq:2}) the summation is only over 
light and heavy neutrinos, respectively. The fields $\chi_k$ and 
$N_k$ satisfy the Majorana condition: 
$\chi_k \xi_k = C ~{\overline{\chi}}_k^T$, 
$N_k {\hat \xi}_k = C ~{\overline{N}}_k^T$,
where C denotes the charge conjugation and $\xi$, ${\hat \xi}$ 
are phase factors.

We assume both outgoing electrons to be
in the $s_{1/2}$ state and consider only 
$0^+_i\rightarrow 0^+_f$ transitions are allowed. 
For the ground state transition restricting ourselves to the
mass mechanism one obtains for the
$0\nu\beta\beta$-decay inverse half-life 
\cite{HS94,DTK85,Verg86,Tom91,PSV96,FS98,SC98},
\begin{equation}
[T_{1/2}^{0\nu}]^{-1} = G_{01} 
|\frac{<m_\nu >}{m_e}M^{light}_{<m_\nu >} + 
\eta_{_N} M^{heavy}_{\eta_{_N}}|^2.
\label{eq:3}   
\end{equation}
The lepton-number non-conserving parameters, i.e. the
effective neutrino mass $<m_\nu >$ and 
$\eta_{_N}$ in Eq. (\ref{eq:3}) are given as follows:
\begin{eqnarray}
<m_\nu > ~ = ~ \sum^{light}_k~ (U^L_{ek})^2 ~ \xi_k ~ m_k, ~ ~~~~~~~
\eta_{_N} ~ = ~ \sum^{heavy}_k~ (U^L_{ek})^2 ~ 
{\hat \xi}_k ~ \frac{m_p}{M_k},
\label{eq:4}   
\end{eqnarray}
with $m_p$ ($m_e$) being the proton (electron) mass. 
$G_{01}$ is the integrated kinematical factor \cite{DTK85,PSV96}. 
The derivation of the nuclear matrix elements associated with the
exchange of light ($M^{light}_{<m_\nu >}$) and heavy 
($M^{heavy}_{\eta_{_N}}$) Majorana neutrinos is outlined in the 
next section. However, Eq. (\ref{eq:3}) applies to any 
intermediate particle.

\subsection{Majoron mechanism}

If the global symmetry associated with lepton number conservation 
is broken spontaneously, the models imply the existence of a physical 
Nambu-Goldstone boson, called Majoron \cite{SV82}, 
which couples to neutrinos:
\begin{equation}
{\cal L}_{\phi\nu\nu} = \sum_{i \le j} 
{\overline{\nu}}_{i} \gamma_5  \nu_j ~(i ~Im~ \phi ) ~P_{ij},
~~~~P_{i,j} = \sum_{\alpha , \beta = e,\mu ,\tau} 
U^R_{i\alpha} ~U^R_{j\beta} ~g_{\alpha \beta},
\label{eq:5}   
\end{equation}
Here, $\nu_i$ denotes both light
$\chi_i$ and heavy $N_i$ Majorana neutrinos. We remind that in analogy 
with Eq. (\ref{eq:5}) there is a unitary transformation for the right 
handed electron neutrino to the mass eigenstates $\chi_k$ and $N_k$:
\begin{equation}
\nu_{e R}=\sum_{k=light} ~U^R_{ek}~\chi_{kR} +
\sum_{k=heavy} ~U^R_{ek}~N_{kR},
\label{eq:6}   
\end{equation}
where $\chi_{kR} = P_R \chi_{k}$ and $N_{kR} = P_R N_{k}$
($P_{R,L}= 1/2(1\pm\gamma_5)$).

Eq. (\ref{eq:5}) leads to Majoron production in 
the $0\nu\beta\beta$-decay ($0\nu\beta\beta\phi$-decay) 
\cite{DTK85,Verg86,FS98,KKS96}.
We are interested in the light neutrino coupling and 
notice that the couplings $U^R_{ek}$ are small
in GUT models where the singlet neutrino is superheavy. 
We restrict our consideration of the $0\nu\beta\beta\phi$-decay  
only to light neutrinos 
( $m_{i,j} \ll q \approx p_F \approx {\cal O}(100 MeV)$ ).
Then  the inverse half-life of the $0\nu\beta\beta\phi$-decay  
 can be written  
\begin{equation}
[T^{\phi 0\nu}_{1/2}]^{-1} = |<g>|^2  |M^{light}_{<m_\nu >}|^2 G_B. 
\label{eq:7}
\end{equation}
Here $<g>$ is the effective Majoron coupling constant
\begin{equation}
<g> = \sum_{i j}^{light} U^L_{ei} U^L_{ej} P_{ij}. 
\label{eq:8}
\end{equation}
The explicit form of the kinematical factor $G_B$
can be found  in Ref. \cite{DTK85}.


\section{The effective transition operator}

Within the impulse approximation the nuclear current  $J^\rho_L$
in Eq. (\ref{eq:1})  expressed with nucleon fields $\Psi$ takes the form
\begin{eqnarray}
J^{\mu \dagger}_L 
=  \overline{\Psi} \tau^+ \left[ g_V(q^2) \gamma^\mu 
- i g_M (q^2) \frac{\sigma^{\mu \nu}}{2 m_p} q_\nu
 - g_A(q^2) \gamma^\mu\gamma_5 + g_P(q^2) q^\mu \gamma_5 \right] \Psi,
\label{eq:9}   
\end{eqnarray}
where $q^\mu = (p-p')_\mu$ is the momentum transferred
from hadrons to leptons ($p$ and $p'$ are four momenta of neutron and 
proton, respectively) and 
$\sigma^{\mu\nu} = (i/2)[\gamma^{\mu}, \gamma^{\nu}]$.
$g_V(q^2)$, $g_M(q^2)$, $g_A(q^2)$ and $g_P(q^2)$ are real functions
of a Lorenz scalar $q^2$. The values of these form factors in the
zero-momentum transfer limit are known as the vector, weak-magnetism,
axial-vector and induced pseudoscalar coupling constants, respectively. 

\subsection{The effective transition operator in momentum space.}

For nuclear structure calculations it is necessary to reduce the nucleon
current to the non-relativistic form. We shall neglect small energy transfers
between nucleons in the non-relativistic expansion. Then the form of the
nucleon current coincides with those in the Breit frame  and we 
arrive at \cite{ERI88},
\begin{equation}
J^\mu(\vec{x})=\sum_{n=1}^A \tau^+_n [g^{\mu 0} J^0({\vec q}^{~2}) +
g^{\mu k} J^k_n({\vec q}^{~2})] \delta(\vec{x}-{\vec{r}}_n),~~~~k=1,2,3,
\label{eq:10}   
\end{equation}
with 
\begin{eqnarray}
J^0({\vec q}^{~2}) = g_V(q^2),~~
{\vec J}_n({\vec q}^{~2}) = g_M({\vec q}^{~2}) 
i \frac{{\vec{\sigma}}_n \times \vec{q}}{2 m_p}+
g_A({\vec q}^{~2})\vec{\sigma}
-g_P({\vec q}^{~2})\frac{\vec{q}~ {\vec{\sigma}}_n \cdot \vec{q}}{2 m_p}.
\label{eq:11}   
\end{eqnarray}
${\vec r}_n$ is the coordinate of the $n$th nucleon.

For the form factors we shall use the following 
parameterization: 
$g_V({\vec q}^{~2}) = {g_V}/{(1+{\vec q}^{~2}/{\Lambda^2_V})^2}$, 
$g_M({\vec q}^{~2})= (\mu_p-\mu_n) g_V({\vec q}^{~2})$, 
$g_A({\vec q}^{~2}) = {g_A}/{(1+{\vec q}^{~2}/{\Lambda^2_A})^2}$ and 
the induced pseudoscalar coupling is given by the
partially conserved axial-vector current hypothesis (PCAC) \cite{towner}
\begin{equation}
g_P({\vec q}^{~2}) = {2 m_p g_A({\vec q}^{~2})}/({{\vec q}^{~2} + m^2_\pi})
(1 - \frac{m_\pi^2}{\Lambda^2_A}),
\label{eq:12}   
\end{equation}
where $g_V = 1$, $g_A = 1.254$, $(\mu_p-\mu_n) = 3.70$, 
$\Lambda^2_V = 0.71 ~(GeV)^2$ \cite{dumb} and $\Lambda_A = 1.09 ~GeV$ 
\cite{towner}. In previous calculations only one 
general cut-off $\Lambda_V = \Lambda_A \approx 0.85$ GeV
was used. In this work 
we take the empirical value of $\Lambda_A $  deduced from the
antineutrino quasielastic reaction 
${\overline{\nu}}_\mu p \rightarrow \mu^+ n $. A larger value of
the cut-off $\Lambda_A $ is expected to increase slightly the values of
the corresponding nuclear matrix elements. It is worth noting that with these
modifications of the nuclear current one gets a new contribution in the neutrino
mass mechanism, namely the tensor contribution. 

As we have already mentioned in the introduction,
momentum dependent terms, in particular the weak-magnetism term, 
have  been considered previously in the $\beta\beta$ decay by 
Tomoda et al. \cite{tom85} and Pantis et al \cite{PSV96},
but in connection with the $\eta$-term. This term is proportional to
the mixing between the vector bosons $W_L$ and $W_R$, which mediate the
left and right handed weak interaction, respectively. 
 They are dominant since, due to their momentum structure, they
can proceed via the s-wave electron wave function, while the standard terms
in this case require p-wave electron wave functions. The pseudoscalar term
is not accompanied by parity change and thus it is not important in the 
extraction of $\eta$. To our knowledge  
this term has not been considered in connection with the
usual light Majorana neutrino mass term of the $0\nu\beta\beta$-decay. 

Under the PCAC hypothesis [see Eq. (\ref{eq:12})] the two body effective
transition operator takes in momentum space the form
\begin{equation}
\Omega = \tau_+ \tau_+ (-h_F + h_{GT} ~ \sigma_{12} - h_T S_{12})
\label{eq:3.4}   
\end{equation}
where the three terms  correspond to Fermi (F) , Gamow-Teller (GT) and Tensor
(T).  One finds that
\begin{equation}
S_{12} = 3({\vec{ \sigma}}_1\cdot \hat{q}
       {\vec{\sigma}}_2 \cdot \hat{q})
      - \sigma_{12}, ~~~ \sigma_{12}=
{\vec{ \sigma}}_1\cdot {\vec{ \sigma}}_2.
\label{eq:3.5}   
\end{equation}
\begin{eqnarray}
h_{F} & = &   ~g^2_V({\vec q}^{~2}) \nonumber \\
h_{GT} ({\vec q}^{~2}) & = & g^2_A({\vec q}^{~2}) ~[ ~1~ -~ 
\frac{2}{3}~ \frac{ {\vec q}^{~2}}{ {\vec q}^{~2} + m^2_\pi } ~+ ~
\frac{1}{3} ~( \frac{ {\vec q}^{~2}}{ {\vec q}^{~2} + m^2_\pi } )^2 ~]
~ +~ \frac{2}{3}~ \frac{g^2_M ({\vec q}^{~2} ) {\vec q}^{~2} }{4 m^2_p }, 
\nonumber \\
h_T ({\vec q}^{~2}) & = & g^2_A({\vec q}^{~2})~ [~ 
\frac{2}{3}~ \frac{ {\vec q}^{~2}}{ {\vec q}^{~2} + m^2_\pi } -
\frac{1}{3}~ ( \frac{ {\vec q}^{~2}}{ {\vec q}^{~2} + m^2_\pi } )^2 ~]~
~+~ \frac{1}{3} ~\frac{g^2_M ({\vec q}^{~2} ) {\vec q}^{~2} }{4 m^2_p },
\label{eq:3.6}    
\end{eqnarray}

 The exact results will depend on the details of the nuclear model, since
the new operators have different momentum (radial) dependence than the
traditional ones and the tensor component is entirely new. We can get a crude
idea of what is happening by taking the above average momentum
 $\langle q \rangle$=100 MeV. Then we find that the GT matrix element 
is reduced by
22$\%$. Then assuming that T matrix element is about half the GT one, we
find that the total reduction is 28$\%$. This will be compared below with the
results of our detailed calculations.

\subsection{The effective transition operator in coordinate space.}

The  nuclear matrix elements entering the half-life formula 
of $0\nu\beta\beta$-decay process take now the form: 
\begin{equation}
M^{I}_{<m_\nu >, \eta_{_N} }=M^{I}_{VV}+M^{I}_{MM}+M^{I}_{AA}+
M^{I}_{AP}+M^{I}_{PP}
\label{eq:13}   
\end{equation}
 with~~ $I=light, ~heavy$.
The partial nuclear matrix elements $M^{I}_{VV}$,  $M^{I}_{MM}$,  
$M^{I}_{AA}$,  $M^{I}_{PP}$ and   $M^{I}_{AP}$ 
have their origin from the vector, the weak-magnetism, 
the axial, the pseudoscalar 
coupling and the interference of the axial-vector and pseudoscalar coupling,
 respectively. They can be expressed in relative 
coordinates by using second quantization. We end up with
formula 
\begin{equation}
M_{type}^I = 
<H_{type-F}^{I}(r_{12})   + 
H_{type-GT}^{I}(r_{12}) {\bf \sigma}_{12} +
H_{type-T}^{I}(r_{12}) { S}_{12})>
\label{eq:14}   
\end{equation}
with $type = VV, MM, AA, PP, AP$ and 
\begin{eqnarray}
{\bf r}_{12} &= &{\bf r}_1-{\bf r}_2, ~~~
r_{12} = |{\bf r}_{12}|, ~~~
\hat{{\bf r}}_{12} = \frac{{\bf r}_{12}}{r_{12}},~\nonumber \\
S_{12} &=& 3({\vec{ \sigma}}_1\cdot \hat{{\bf r}}_{12})
       ({\vec{\sigma}}_2 \cdot \hat{{\bf r}}_{12})
      - \sigma_{12}, ~~~ \sigma_{12}=
{\vec{ \sigma}}_1\cdot {\vec{ \sigma}}_2.
\label{eq:15}   
\end{eqnarray}
${\bf r}_1$ and ${\bf r}_2$ are coordinates of the beta decaying nucleons.
The form of the matrix element 
$<{\cal O}(1,2)>$ within the pn-RQRPA will be presented in the next section.

The light and heavy neutrino-exchange potentials 
$H^{light,heavy}_{type-K}(r_{12})$
($K = F, GT, T$) have the following forms,
\begin{eqnarray}   
H^{light}_{type-K}(r_{12})&=&\frac{2}{\pi g_A^2}
\frac{R}{r_{12}} \int_{0}^{\infty} 
\frac{\sin(qr_{12})} {q+E^m_{J} - (E^i_{g.s.} + E^f_{g.s.})/2}
h_{type-K}(q^2)~d{q}, ~~
\label{eq:16}   \\
H^{heavy}_{type-K}(r_{12})&=&\frac{1}{m_p m_e}\frac{2}{\pi g^2_A}
\frac{R}{r_{12}} \int_{0}^{\infty} 
{\sin(qr_{12})}h_{type-K}(q^2)~ q~ d{q}
\label{eq:17}   
\end{eqnarray}
Here, $E^i_{g.s.}$, $E^f_{g.s.}$ and 
$E^{m}_{J}$ are respectively the energies of the 
initial, final and intermediate nuclear states. 
$R = r_0 A^{1/3}$ is the mean nuclear radius, 
with $r_0 = 1.1~ fm$. 
The relevant couplings are: 
\begin{eqnarray}
h_{VV}({\vec q}^{~2} ) &=& - g^2_{V} ({\vec q}^{~2})\nonumber \\
h_{MM-GT}({\vec q}^{~2}) &=& 
\frac{2}{3} \frac{g^2_M ({\vec q}^{~2}) {\vec q}^{~2} }{4m^2_p},~~~~~~~
h_{MM-T}({\vec q}^{~2}) = 
\frac{1}{3} \frac{g^2_M ({\vec q}^{~2}) {\vec q}^{~2}}{4m^2_p}, 
\nonumber \\ 
h_{AA-GT}({\vec q}^{~2}) &=& g^2_A ({\vec q}^{~2}), 
\nonumber \\
h_{PP-GT}({\vec q}^{~2}) &=& 
\frac{1}{3}\frac{g^2_P ({\vec q}^{~2}) {\vec q}^{~4}}{4 m^2_p},~~~~~~~~~~~
h_{PP-T}(q^2) = - \frac{1}{3}\frac{g^2_P ({\vec q}^{~2}) 
{\vec q}^{~4}}{4 m^2_p},
\nonumber \\ 
h_{AP-GT}({\vec q}^{~2}) &=& -\frac{2}{3}\frac{g_A ({\vec q}^{~2}) 
g_P ({\vec q}^{~2}) {\vec q}^{~2}}{2 m_p},~~
h_{AP-T}({\vec q}^{~2}) = \frac{2}{3}\frac{g_A ({\vec q}^{~2}) 
g_P ({\vec q}^{~2}) {\vec q}^{~2}}{2 m_p}.
\label{eq:18}   
\end{eqnarray}
The tensor form factor includes a sign change 
going from momentum to coordinate space.

The full matrix element is of the form:
\begin{equation}
M^{light}_{<m_\nu >} = - \frac{M_F^{light}}{g^2_A} + M_{GT}^{light}
+ M_{T}^{light}.
\label{eq:20}    
\end{equation}
We see that the Fermi component is unchanged, the Gamow-Teller is
modified and  the tensor component appeared due to the new terms. 


\section{Nuclear structure ingredients}

 As we have mentioned above, we would like to evaluate the changes in
the $0\nu\beta\beta$-decay nuclear matrix elements due to the modifications
of the nuclear current introduced above, relevant for the neutrino mass
mechanism. It is clear that the $0\nu\beta\beta$-decay 
is a second order process in the
weak interaction and, thus, the corresponding nuclear matrix 
elements require the summation over a complete set
of intermediate nuclear states. Even though the construction of these states
is not needed, a closure approximation with a reasonable average energy 
denominator is very  accurate \cite{Ver90,PV90,FKPV91}, the initial
and final states
of the nuclear systems, which can undergo double beta decay, are not easy to
construct, since these nuclei are far removed from closed shells. Thus the 
introduction of additional approximations is necessary.

Thus at this point we will reduce the computational difficulty in evaluating
the effects of the above mentioned modifications, by using
the proton-neutron Quasiparticle Random Phase Approximation
(pn-QRPA) \cite{VZ86,CAT87,MBK88,RFSK91}, 
which is an approximation to solve the nuclear  
many-body problem. Admittedly the intermediate states must be explicitly
constructed in this case, but in this scheme it is a a simple matter to
include even a large number of such states, if necessary.

 The crucial simplifying point of the QRPA is the quasiboson 
approximation (QBA) assuming nuclear excitations to be harmonic.
It  leads to a violation of the Pauli exclusion principle. 
The drawback of this approximation is that
with increasing strength of the nucleon-nucleon force in the 
particle-particle channel  
the QRPA overestimates  the ground state
correlations and the QRPA solution collapses
\cite{TS95,SSF96,SSVP97,SPF98}.

The renormalized QRPA (RQRPA) \cite{TS95,SSF96}
overcomes this difficulty by taking 
into account the Pauli exclusion principle in a more proper
way by using the renormalized QBA. In this  paper
we calculate the nuclear matrix elements of the $0\nu\beta\beta$-decay 
within the proton-neutron RQRPA (pn-RQRPA) \cite{TS95,SSF96}, 
which is an extension of the pn-QRPA by incorporating the Pauli
exclusion principle for fermion pairs. For studying the relative importance
of the new induced  terms, which is the main thrust of
our paper,  the inclusion of other refinements like p-n pairing, which is 
computationally very involved, is not essential.

Furthermore p-n pairing is rigorously incorporated in the BCS
ansatz only for the T=1 states, while T=0 p-n pairing effects are implicitly 
taken into account. This procedure seems to avoid the collapse of the QRPA 
within the physical
region of the Hamiltonian but we are not sure whether in some cases it does not 
produce more ground state correlations which can lead to strong cancellations in the
matrix element. This might have been the case of $^{100}Mo$ in our earlier work \cite{PSV96}
in which one should have expected a minor influence of p-n pairing. In fact if the
collapse of the QRPA reflects a nearby phase transition \cite{VOG98}, i.e.,  a change of the   
ground state from being dominated by T=1 pair correlations to being dominated by T=0 pair
correlations, further work needs to been done to be sure about the p-n pairing.
On  the other hand  the renormalized  solution does not lead to a 
collapse of the QRPA for  physical values of the proton-neutron interaction
strength and is tractable from the computational point of view.

The pn-RQRPA excited states $|m, JM>$ are of the form \cite{TS95,SSF96}
\begin{eqnarray}
|J^\pi M m\rangle  &=& Q^{m\dagger}_{JM^\pi}|0^+_{RPA}\rangle  \nonumber\\
& = & \sum_{pn} [ X^m_{(pn, J^\pi)} A^\dagger(pn, JM)
+ Y^m_{(pn, J^\pi)}\tilde{A}(pn, JM)] |0^+_{RPA}\rangle ,
\label{eq:21}
\end{eqnarray}
where $X^m_{(pn, J^\pi)}$ and $Y^m_{(pn, J^\pi)}$ are 
free variational amplitudes, respectively,  and
\begin{equation}
A^\dagger(pn, JM)  =  \sum^{}_{m_p , m_n }
C^{J M}_{j_p m_p j_n m_n } a^\dagger_{p m_p} a^\dagger_{n m_n}, 
~~~~~
\tilde{A}(pn, JM)  =   (-1)^{J-M}{A}(pn, J -M). 
\label{eq:22}
\end{equation}
Here, $a^{+}_{\tau m_\tau}$ ($a^{}_{\tau m_\tau}$, $\tau = p,n$) 
is the quasiparticle creation (annihilation) operator for 
spherical shell model states, which is related to 
the particle creation and annihilation 
($c^{+}_{\tau m_\tau}$ and $c^{}_{\tau m_\tau}$, $\tau = p,n$) 
operators by the 
Bogoliubov-Valatin transformation: 
\begin{equation}  
\left( \matrix{ c^{+}_{\tau m_{\tau} } \cr
 {\tilde{c}}_{\tau  m_{\tau} } 
}\right) = \left( \matrix{ 
u_{\tau} & -v_{\tau} \cr 
v_{\tau} & u_{\tau} 
}\right)
\left( \matrix{ a^{+}_{\tau m_{\tau}} \cr
{\tilde{a}}_{\tau m_{\tau}} 
}\right).
\label{eq:23}
\end{equation} 
where the tilde $\sim$ indicates  time reversal 
($\tilde{a}_{\tau {m}_{\tau}}$ = 
$(-1)^{j_{\tau} - m_{\tau}}a^{}_{\tau -m_{\tau}}$). 
The label $\tau$ designates quantum numbers 
$n^{}_\tau, l^{}_{\tau}, j^{}_{\tau}$. 
The occupation amplitudes $u$ and $v$ and the single quasiparticle
energies $E_\tau$ are obtained by solving the BCS equation.

In the pn-RQRPA the commutator of two-bifermion operators
fulfill the following relation (renormalized QBA)
\begin{eqnarray}
&& \big<0^+_{RPA}\big|\big
[A^{} (pn, JM), A^+(p'n', JM)\big]
\big|0^+_{RPA}\big> = \delta_{pp'}\delta_{nn'}\times
\nonumber \\ 
\lefteqn{\underbrace{
\Big\{1
\,-\,\frac{1}{\hat{\jmath}_{l}}
<0^+_{RPA}|[a^+_{p}{\tilde{a}}_{p}]_{00}|0^+_{RPA}>
\,-\,\frac{1}{\hat{\jmath}_{k}}
<0^+_{RPA}|[a^+_{n}{\tilde{a}}_{n}]_{00}|0^+_{RPA}>
\Big\}
}_{
\displaystyle {\cal D}_{pn, J^\pi}
},} &&
\nonumber \\
\label{eq:24}
\end{eqnarray}
with  $\hat{\jmath}_p=\sqrt{2j_p+1}$.
If we replace $|0^+_{RPA}>$ in Eq. (\ref{eq:24}) with the uncorrelated
BCS ground state, we obtain the quasiboson approximation (i.e. 
${\cal D}_{pn, J^\pi}=1$), which  assumes that  pairs of quasiparticles 
obey the commutation relations of bosons. We note that it is convenient
to introduce amplitudes 
\begin{equation}
{\overline{X}}^m_{(pn, J^\pi)}  =  {\cal D}^{1/2}_{pn, J^\pi}~ 
X^m_{(pn, J^\pi)}, ~~~~~
{\overline{Y}}^m_{(pn, J^\pi)}  =  {\cal D}^{1/2}_{pn, J^\pi}~ 
Y^m_{(pn, J^\pi)},
\label{eq:25}   
\end{equation}
which are orthonormalized in the usual way 
\begin{equation}
\delta_{m m'} = 
\sum_{p n}
\big(
\overline{X}^{m}_{(pn, J^\pi)}
\overline{X}^{m'}_{(pn, J^\pi)}-
\overline{Y}^{m}_{(pn, J^\pi)}
\overline{Y}^{m'}_{(pn, J^\pi)} \big) .
\label{eq:26}   
\end{equation}

One can show that 
\begin{equation}
\label{eq:27}
\frac{<J^\pi m_i\parallel [c^+_{p}{\tilde{c}}_{n}]_J\parallel 0^+_i>}
{\sqrt{2J+1}} = 
(u_{p}^{(i)} v_{n}^{(i)} {\overline{X}}^{m_i}_{(pn, J^\pi)}
+v_{p}^{(i)} u_{n}^{(i)} {\overline{Y}}^{m_i}_{(pn, J^\pi)})
\sqrt{{\cal D}^{(i)}_{pn,  J^\pi}}, 
\end{equation}
\begin{equation}
\frac{<0_f^+\parallel\widetilde{ [c^+_{p}{\tilde{c}}_{n}]_J}
\parallel J^\pi m_f> }{\sqrt{2J+1} } =
(v_{p}^{(f)} u_{n}^{(f)} 
{\overline{X}}^{m_f}_{(pn, J^\pi)}
+u_{p}^{(f)} v_{n}^{(f)} 
{\overline{Y}}^{m_f}_{(pn, J^\pi)})
\sqrt{{\cal D}^{(f)}_{pn, J^\pi}}.
\label{eq:28}
\end{equation}
The index i (f) indicates that the quasiparticles and the excited
states of the nucleus are defined with respect to the initial (final)
nuclear ground state $|0^+_i>$ ($|0^+_f>$). The forward- and
backward- amplitudes ${\overline{X}}^{m_i}_{(pn, J^\pi)}$ and
${\overline{Y}}^{m_i}_{(pn, J^\pi)}$ and the energies of the 
excited states $\Omega^{m_i}_{J^\pi} = E^{m_i}_{J^\pi} -E^i_{g.s.}$
are obtained by solving the non-linear set of RQRPA equations for
the initial nucleus (A,Z) \cite{TS95,SSF96}. By performing  the
RQRPA diagonalization for the final nucleus (A,Z=2) we obtain the 
amplitudes ${\overline{X}}^{m_f}_{(pn, J^\pi)}$ and
${\overline{Y}}^{m_f}_{(pn, J^\pi)}$ and the eigenenergies 
$\Omega^{m_f}_{J^\pi} = E^{m_f}_{J^\pi} -E^f_{g.s.}$ of the
RQRPA state $|J^\pi m_f \rangle $.

Within the pn-RQRPA the $0\nu\beta\beta$-decay
matrix elements given in Eqs.\ (\ref{eq:14}) take the
following form:
\begin{eqnarray}
\label{eq:29}
M^I_{type} =
\sum_{{p n p' n' } \atop {J^{\pi}
m_i m_f {\cal J}  }}
~(-)^{j_{n}+j_{p'}+J+{\cal J}}(2{\cal J}+1)
\left\{
\matrix{
j_p &j_n &J \cr
j_{n'}&j_{p'}&{\cal J}}
\right\}\times~~~~~~~~\nonumber \\
<p(1), p'(2);{\cal J}|f(r_{12})\tau_1^+ \tau_2^+ 
{\cal O}_{type}^I (12)
f(r_{12})|n(1) ,n'(2);{\cal J}>\times ~~~~
\nonumber \\
< 0_f^+ \parallel
\widetilde{[c^+_{p'}{\tilde{c}}_{n'}]_J} \parallel J^\pi m_f>
<J^\pi m_f|J^\pi m_i>
<J^\pi m_i \parallel [c^+_{p}{\tilde{c}}_{n}]_J \parallel
0^+_i >.
\end{eqnarray}
Here, ${\cal O}_{type}^I (12)$ 
represents the coordinate and spin dependent part of the
two body transition operators of the $0\nu\beta\beta$-decay
nuclear matrix elements in Eq.  (\ref{eq:29}) 
\begin{equation}
{\cal O}_{type}^I (12) = 
H_{type-F}^{I}(r_{12})   + 
H_{type-GT}^{I}(r_{12}) {\bf \sigma}_{12} +
H_{type-T}^{I}(r_{12}) {\bf S}_{12}.
\label{eq:30}
\end{equation}
The short-range correlations
between the two interacting protons ($p(1)$ and $p'(2)$)
and neutrons ($n(1)$ and $n'(2)$) are taken into account by
the correlation function $f(r_{12})$ in the non-antisymmetrized two-body 
matrix element in Eq. (\ref{eq:29}). $f(r_{12})$ is given as 
follows:
\begin{equation}
\label{eq:31}
f(r_{12})=1-e^{-\alpha r^2_{12} }(1-b r^2_{12}) \quad \mbox{with} \quad
\alpha=1.1~ \mbox{fm}^2 \quad \mbox{and} \quad  b=0.68 ~\mbox{fm}^2.
\end{equation}

For the overlap matrix of intermediate nuclear states generated from the 
initial and final ground states we write \cite{SPF98}:
\begin{eqnarray}
\label{eq:32}
< J^\pi M m_f | J^\pi M m_i > &\approx & 
\sum_{p n}
\big(
\overline{X}^{m_{i}^{}}_{(pn, J^\pi)}
\overline{X}^{m_{f}^{}}_{(pn, J^\pi)}-
\overline{Y}^{m_{i}^{}}_{(pn, J^\pi)}
\overline{Y}^{m_{f}^{}}_{(pn, J^\pi)} 
\big)\times \nonumber \\ 
&& ~(u^{(i)}_{p}u^{(f)}_{p} +v^{(i)}_{p}v^{(f)}_{p}) 
(u^{(i)}_{n}u^{(f)}_{n} +v^{(i)}_{n}v^{(f)}_{n}). 
\end{eqnarray}


\section{Calculation, Discussion and Outlook}

 In order to test he importance of the new momentum dependent terms in the
nucleon current,
we applied the pn-RQRPA to calculate the $0\nu\beta\beta$-decay
of the $A =$ $76$, $82$, $96$, $100$, $116$, $128$, $130$, $136$ and 
$150$ systems. To this end the considered single-particle
model spaces both for protons and neutrons have been as follows:
i) For A=76, 82  the model space consists of the full
$2-4\hbar\omega$ major oscillator shells.
ii) For A=96, 100, 116 we added to the previous model space
$1f_{5/2}$, $1f_{7/2}$,
$0h_{9/2}$ and $0h_{11/2}$ levels.
iii) For A=128, 130, 136 the model space
comprises the full $2-5\hbar\omega$ major shells.
iv) For A=150 the model space extends over the full $2-5\hbar\omega$ shells
plus the $0i_{11/2}$ and $0i_{13/2}$ levels.

The single particle
energies were  obtained by using a  Coulomb corrected Woods Saxon
potential.  The interaction employed was the Brueckner G-matrix which is a
solution of the Bethe-Goldstone equation with the Bonn
one-boson exchange potential. Since the model space considered is finite, 
the pairing interactions have been adjusted to fit 
the empirical pairing gaps according to \cite{cheoun}. In addition, 
we renormalize the  particle-particle and particle-hole channels of 
the G-matrix interaction of the nuclear Hamiltonian $H$ by introducing
the parameters $g_{pp}$ and $g_{ph}$, respectively. The nuclear matrix 
elements listed in Tables \ref{table.1} and \ref{table.2} have been obtained 
for $g_{ph}=0.8$ and $g_{pp}=1.0$. With respect to the $g_{pp}$ we
wish to make the following statement: Our numerical results do not show 
significant variations (do not exceed $20~\%$) in 
the physical region of $g_{pp}$ ($0.8 \le g_{pp} \le 1.2$).

A detailed study of Fermi, Gamow-Teller and Tensor contribution to 
the full nuclear matrix element $M^{light}_{<m_\nu >}$ in 
Eq. (\ref{eq:20}) for the two representative $0\nu\beta\beta$-decay nuclei
$^{76}Ge$ and $^{130}Te$ is presented in Table \ref{table.1}. One
notices significant additional contributions to GT (AA and PP)
and tensor (AA and PP) nuclear matrix elements coming from
the induced current terms. By glancing at the 
Table \ref{table.1} we also see that Fermi and GT matrix elements are
strongly suppressed due to the nucleon short-range correlations.

The relative importance of the different contributions to 
the nuclear matrix elements $M^{light}_{<m_\nu >}$ 
and $M^{heavy}_{\eta_{_N}}$ [see Eq. (\ref{eq:13})] is displayed in 
Fig. \ref{fig.1} for the A=76 and 130 nuclear
systems. As expected from our discussion in Section III we see that 
for light neutrinos the pseudoscalar term, in particular the AP
contribution, is important. It is in fact as important as the
usual vector contribution but in  the opposite direction  (see Table 
\ref{table.2}). Our calculations verify our above estimate, i.e. the new 
terms in the hadronic current, and in particular the induced pseudoscalar term,
tend to increase the average neutrino mass $|<m_\nu >|$ and 
average Majoron coupling $<g>$ from experiment by about $30~\%$. 
They are much more important in the exchange of heavy neutrinos
leading the suppression of $M^{heavy}_{\eta_{_N}}$ by about 
factor of 3-6 (see Table \ref{table.2}). 
The contributions from previously neglected 
$M^{heavy}_{MM}$ and $M^{heavy}_{AP}$ to 
$M^{heavy}_{\eta_{_N}}$ are much more important as that from 
$M^{heavy}_{VV}$. A large value of  $M^{heavy}_{MM}$, which 
has its origin in weak-magnetism, indicates that 
the average neutrino momentum is large, i.e. about the order of 
magnitude of the nucleon mass.

We present in Fig. \ref{fig.2} the nuclear matrix elements 
$M^{light}_{<m_\nu >}$ and $M^{heavy}_{\eta_{_N}}$ calculated within
pn-RQRPA for the A = 76, 82, 96, 100, 116, 128, 130, 136 and 150 nuclear
systems. We see that the inclusion of the induced pseudoscalar interaction
and of weak-magnetism
in the calculation results in considerably smaller nuclear
matrix elements for all nuclear systems. The numerical
values of $M^{light}_{<m_\nu >}$ and $M^{heavy}_{\eta_{_N}}$ can be found
in Table  \ref{table.2}. The largest matrix elements 
of $M^{light}_{<m_\nu >}$  for A=150, 100 and 76 are
3.33, 3.21 and 2.80, respectively. For A=150, 76 and 82
the largest values of   
$M^{heavy}_{\eta_{_N}}$ are 35.6, 32.6 and 30.0, respectively.
We notice that the A=136 system has the smallest 
nuclear matrix elements: $M^{light}_{<m_\nu >}=0.66$ and 
$M^{heavy}_{\eta_{_N}}=14.1$. We suppose that it is connected with fact 
that $^{136}Xe$ is a closed shell nucleus for neutrons (N=82). The 
sharp Fermi level for neutrons  yield smaller $0\nu\beta\beta$-decay
matrix elements.  
We note that in calculating the matrix elements involving the 
exchange of heavy neutrinos, the treatment of the short-range 
repulsion and nucleon finite size is crucial. We have found that 
the consideration of short range correlation effects reduces
the values of $M^{heavy}_{\eta_{_N}}$ by about factor of 20-30.
As we mentioned already the nucleon finite size has been 
taken into account through the phenomenological formfactors and
the PCAC hypothesis.  However, the choice of the formfactor can 
influence the results significantly as it was manifested in Ref.
\cite{SEIL92} performing the calculations with both phenomenological
and quark confinement model formfactors. 

The limits deduced for lepton-number violating parameters depend 
on the values of nuclear matrix element, of the kinematical factor and
of the current experimental limit for a given isotope
[see Eqs. (\ref{eq:3}) and (\ref{eq:7})]. It is expected that 
the experimental constraints on the half-life of the $0\nu\beta\beta$-decay 
are expected to be more stringent in future. Thus there is useful
to introduce sensitivity parameters for a given isotope to the effective 
light and heavy Majorana neutrino mass and Majoron signals, which depend 
only on the characteristics of a given nuclear system.
There are as follows:
\begin{eqnarray}
\label{eq:33}
\zeta_{<m_\nu >} (Y) & = &
10^{7}~ |M^{light}_{<m_\nu >}|~ 
\sqrt{{G_{01}}~ {year}},~~~~~~
\frac{<m_\nu >}{m_e} \leq \frac{10^{-5}}{\zeta_{<m_\nu >}}
\sqrt{\frac{10^{24}~years}{T^{0\nu -exp}_{1/2}}}, \nonumber \\
\zeta_{\eta_{_N}} (Y) & = & 
10^{6}~ |M^{heavy}_{<m_\nu >}|~ 
\sqrt{{G_{01}}~{year}},~~~~~~
\eta_{_N} \leq \frac{10^{-6}}{\zeta_{\eta_{_N}}}
\sqrt{\frac{10^{24}~years}{T^{0\nu -exp}_{1/2}}}, \nonumber \\
\zeta_{<g>} (Y) & = & 
10^{8}~ |M^{light}_{<m_\nu >}|~ 
\sqrt{{G_{B}}~{year}},~~~~~~
<g> \leq \frac{10^{-4}}{\zeta_{<g>}}
\sqrt{\frac{10^{24}~years}{T^{0\nu -exp}_{1/2}}}.
\end{eqnarray}
The normalization of $10^{24}$ years was chosen so that the 
$\zeta$'s are of order unity. 
The numerical values of these parameters for the nuclear systems 
considered in this work are listed in Table \ref{table.2} and can be used
in predicting desired limits on lepton number violation
with changing experimental data (see the above expressions). 
These characteristics estimate 
also prospects for searches of light and heavy Majorana neutrinos and 
of the Majoron. The larger values of these 
parameters determine those $0\nu\beta\beta$-decay isotopes,
which are the most promising candidates for searching for the 
corresponding lepton number violating signal. They give
information on the requirements on a 
$0\nu\beta\beta$-detector. 

Fig.\ref{fig.2} shows that 
the most sensitive isotope to all three studied lepton number
violating parameters is $^{150}Nd$. It is mostly due to the large phase space
integral and to some extent also to the larger nuclear matrix elements. We 
caution the reader here that our matrix elements are not model independent in 
the sense that  all the nuclei
considered in this work are treated as spherical . The nucleus $^{150}Nd$,
is  deformed and our results may not be the same quantitatively,
were we to perform calculations taking into account effects of nuclear
deformation.

 The purpose of this work, however, is to study the effects of the induced 
currents. It is thus reasonable to do this by comparing calculations within the 
same model. This means that the model itself will play a minor  role, if any, 
in the investigation of such effects. We are thus satisfied that for light
neutrinos the effect is almost the same throughout the periodic table.
In addition, as we see from Fig. \ref{fig.2}, the
inclusion of weak-magnetism and induced 
pseudoscalar coupling in our calculations
leads to a significant reduction of the parameters 
of $\zeta_{<m_\nu >} (Y)$, $\zeta_{\eta_{_N}} (Y)$ and 
$\zeta_{<g>} (Y)$.

The present experimental situation in terms of the accessible 
half-life and the corresponding upper limit on $<m_\nu >$,
$\eta_{_N}$ and $<g>$ is given in Table \ref{table.3}. 
Thus, the most restrictive limits are as follows:
\begin{eqnarray}
<m_\nu >^{best} ~&<& ~~ 0.62~ eV, ~~~~~~~~~~~~~~[^{76}Ge,~~ Ref.~ 38]
\nonumber \\
\eta_{_N}^{best}~ &<&  ~~1.0\times 10^{-7},
~~~~~~~~~[^{76}Ge, ~~Ref.~ 38],
\nonumber \\
<g>^{best} ~ &<&  ~~6.9\times 10^{-5}.~~~~~~~~~~[^{128}Te,~~Ref.~ 43].
\label{vio.1}
\end{eqnarray}
The sensitivity  of different experiments to $<m_\nu >$,  $<\eta_{_N}>$ and
$<g> $ is drawn in Fig. \ref{fig.3}. Currently, the Heidelberg-Moscow
experiment \cite{hdmo97} offers the most stringent limit for 
effective light and heavy Majorana neutrino mass and the $^{128}Te$  
experiment \cite{bern92} for effective Majoron coupling constant. 
By  assuming  $<m_\nu > = <m_\nu >^{best}$,  
$\eta_{_N} = \eta_{_N}^{best}$ and $<g> = <g>^{best}$ in Eqs. 
(\ref{eq:3}) and (\ref{eq:7}) 
we calculated half-lifes of the $0\nu\beta\beta$-decay
$T^{exp-0\nu}_{1/2}$($<m_\nu >^{best}$), 
$T^{exp-0\nu}_{1/2}$($\eta^{best}_{_N}$) and 
$T^{exp-0\nu\phi}_{1/2}$($<g>^{best}$) for nuclear systems of interest
using specific mechanisms with the " best " parameters. 
These corresponding numerical values are
listed in Table \ref{table.3} and shown by open bars in Fig.
\ref{fig.3}. Since the quantities $<m_\nu >$, $\eta_{_N}$, $<g>$
depend only on particle theory parameters these quantities 
indicate the experimental half-life limit for a
given isotope, which the relevant experiments should 
reach in order to extract the best
present bound on the corresponding lepton number violating 
parameter from their data. Some of them have a long way to go to
reach the $^{76}Ge$ target limit.  

At present  most attention is paid to the light Majorana
neutrino mass because of the experimental indications for 
oscillations  of solar (Homestake \cite{cle95}, Kamiokande \cite{hir91}, 
Gallex \cite{gal95} and SAGE \cite{abd94}), atmospheric 
(Kamiokande \cite{fuk94}, IMB \cite{bec95} and 
Soudan \cite{goo96}, Super-Kamiokande experiments \cite{Fuk98})
and terrestrial neutrinos (LSND experiment \cite{ath95}).
One can use the constraints imposed by the results  of neutrino 
oscillation experiments on $<m_\nu >$. The predictions differ 
from each other due the different input and structure of the 
neutrino  mixing matrix and in particular the assumed Majorana
condition phases. Bilenky et al \cite{biln98} have shown 
that in a general scheme with three light Majorana neutrinos and 
mass hierarchy  $|<m_\nu >|$ is smaller than $10^{-2}$ eV. 
In another study  outlined in Ref.\cite{BS98} the authors end up
with $|<m_\nu >| \approx 0.14$ eV. Bednyakov, Faessler and Kovalenko
considered neutrino oscillations within the minimal supersymmetric
standard model with R-parity breaking. They showed that 
Super-Kamiokande atmospheric data are compatible with
$|<m_\nu >| \le 0.8\times 10^{-2}$ eV \cite{BFK98}. One sees that, the
current limit on $<m_\nu >$
in (\ref{vio.1}) is quite a bit higher than the neutrino oscillation
data.

There is 
a new experimental proposal for measurement of the $0\nu\beta\beta$-decay
of $^{76}Ge$, which intents to use 1 ton (in an extended version 10 tons)
of enriched $^{76}Ge$ and to reach  the half-life limit
$T^{0\nu -exp}_{1/2} \geq 5.8\times 10^{27}$ and  
$T^{0\nu -exp}_{1/2} \geq 6.4\times 10^{28}$ after one ( and 10 years)
of measurements, respectively. From these half-life values one can 
deduce [see Eq. (\ref{eq:33}) and Table \ref{table.2}] the possible
future limits on the effective light neutrino mass 
$2.7\times 10^{-2}$ eV and $8.1\times 10^{-3}$ eV, respectively. 

 By comparing the above limits with those advocated by the neutrino oscillation
phenomenology we conclude that the GENIUS experiment will be able to reach 
similar limits, provided, of course, that the neutrinos are Majorana particles.
We remind that there is also a plethora of other $0\nu\beta\beta$-decay
mechanisms  predicted by GUT's and SUSY. One can show,
however, that their presence implies that the neutrinos are massive 
Majorana particles even if the mass mechanism is not 
dominant \cite{SVa82,kov97}. Certainly, the 
experimental detection of the $0\nu\beta\beta$-decay process would be
a  major achievement with important implications 
on the field of particle and nuclear physics
as well as on cosmology.


\section{Conclusions}

The contributions coming from the induced currents at the nucleon level, 
such as the weak-magnetism and induced 
pseudoscalar coupling  on the mass mechanism
for the $0\nu\beta\beta$-decay transitions has been studied.  The needed
nuclear matrix elements, associated with 
the light and heavy Majorana neutrinos as well as the Majoron  emission
mechanisms, have been obtained in the context of pn-RQRPA, which is known to
produce results more reliable than the  standard QRPA. 
Our results are shown in Figs. \ref{fig.1}-\ref{fig.3} 
and listed in Tables \ref{table.1}-\ref{table.3}.
One can see that the modification of the nuclear current due to the 
weak-magnetism and  induced pseudoscalar coupling is important and 
results in considerable reductions of the $0\nu\beta\beta$-decay 
matrix elements. For the light neutrino exchange this reduction 
amounts to about $20-30\%$ for all nuclei considered. 
The reductions for the heavy neutrino exchange are even more 
significant with factors ranging from 4 to 6. 

The derived upper 
limits on $<m_\nu >$, $\eta_{_N}$ and $<g>$ from the current 
experimental limits of the $0\nu\beta\beta$-decay 
lifetime for A = 76, 82, 96, 100, 116, 128, 130, 136 and 150  
are listed in Table \ref{table.3}. 
This makes the extracted limits  of the lepton number violating parameters 
less stringent yielding 
$<m_\nu >^{best} \le 0.62~ eV$,  
$\eta_{_N}^{best}~ \le  ~~1.0\times 10^{-7}$, 
$<g>^{best} ~ \le  ~6.9\times 10^{-5}$ deduced from the 
$^{76}Ge$ \cite{hdmo97} and $^{128}Te$  \cite{bern92} data.
Further, we introduced and evaluated useful 
sensitivity parameters  for various 
lepton number violating signals for some nuclei of interest, which might 
be helpful in planning future $0\nu\beta\beta$-decay experiments. 
 
The value of $\eta_{_N}$ extracted is, of course, associated with heavy 
Majorana neutrino. It can, however, be applicable in other
processes involving the exchange of heavy particles, provided that 
the momentum structure of the relevant operators is not very 
different from that in Eqs. (\ref{eq:16})-(\ref{eq:18}).

 Admittedly there is a rather large spread between the 
calculated values of nuclear matrix elements within different 
nuclear theories [see e.g. recent review articles \cite{FS98,SC98}],
which could be considered as a measure of the theoretical uncertainty.
Between some of them there is no objective way to judge which 
calculation is correct. However, one can argue that the RQRPA method
offers more reliable results than the QRPA primarily because of the
collapse of the QRPA solution and the strong  sensitivity
of the QRPA results to the strength of particle-particle force. 
The only advantage of the QRPA over RQRPA is that it fulfills the
Ikeda sum rule. However, the meaning of this fact is questionable because 
it is so close to the collapse of the QRPA, where the obtained solution 
is far from being realistic. In the present calculations we are using the 
pn-RQRPA and we take into account
 also additional nucleon currents effects. Thus we consider
the results of this paper more reliable in respect to pn-QRPA results
of our and other groups. 
In this work we did not deal with the problem of the proton-neutron
pairing. The effects of proton-neutron pairing within the 
renormalized QRPA  have been discussed for some nuclei of
interest in Refs. \cite{SSVP97}.

Be that as it may, we find that in the case of light neutrino the momentum 
dependent terms in the nucleon current cause a more or less uniform  reduction 
of the nuclear matrix elements by approximately $30\%$ throughout
the periodic table. We expect a similar reduction in almost any nuclear model.
This will cause a corresponding increase of the extracted values for the
neutrino mass.

We thus conclude that, with the best nuclear physics input, the extracted 
average neutrino mass is low, but quite a bit higher than that deduced from
the present neutrino oscillation experiments. It will reach, however, similar 
levels, if the planned experiments reach the sensitivity aimed at by  
the GENIUS experiment \cite{HK97}. In any case the neutrino oscillation data 
can neither set the absolute scale of the mass nor decide whether the neutrino
is a Majorana particle. The latter issue can be decided only by the
$0\nu\beta\beta$-decay.

\acknowledgments

One of us (JDV) would like to express his appreciation to the
Humboldt Foundation for their award.  JDV and F\v S thank the 
Institute of Theoretical Physics at the University of T\" ubingen
for its hospitality.


\widetext

\begin{table}[t]
\caption{The Fermi, Gamow-Teller and Tensor 
nuclear matrix elements for the light Majorana neutrino
exchange of the $0\nu\beta\beta$-decay of
$^{76}Ge$ and $^{130}Te$ with and 
without consideration of short-range correlations (s.r.c.).
}
\label{table.1}
\begin{tabular}{lrrrrrrrrr}
transition & s.r.c.  &\multicolumn{3}{c}{ Gamow-Teller} & 
  \multicolumn{2}{c}{Tensor} & $M^{light}_F$ & 
  $M^{light}_{GT}$ & $M^{light}_T$ \\
  &   & AA & AP & PP & AP & PP & & & \\ \hline
$^{76}Ge \rightarrow ^{76}Se$ & without & 
 5.132 & -1.392 & 0.302 & -0.243 & 0.054 & -2.059 & 4.042  & -0.188 \\
  & with & 
 2.797 & -0.790 & 0.176 & -0.246 & 0.055 & -1.261 & 2.183 & -0.190 \\
$^{130}Te \rightarrow ^{130}Xe $ & without &
 4.158 & -1.173 & 0.258 & -0.329 & 0.074 & -1.837 & 3.243 & -0.255 \\
  & with & 
 1.841 & -0.578 & 0.134 & -0.333 & 0.075 & -1.033 & 1.397 & -0.258 \\
\end{tabular}
\end{table}

\begin{table}[t]
\caption{Nuclear matrix elements for the light and heavy 
Majorana neutrino exchange modes of the $0\nu\beta\beta$-decay for the
nuclei studied in this work calculated within
the renormalized pn-QRPA. $G_{01}$ and $G_{B}$
are the integrated kinematical factors for the 
$0^+ \rightarrow 0^+$ transition.
$\zeta_{<m_\nu >} (Y)$, $\zeta_{\eta_N} (Y)$ and $\zeta_{<g>} (Y)$ 
denote according to Eq. (34) the sensitivity of
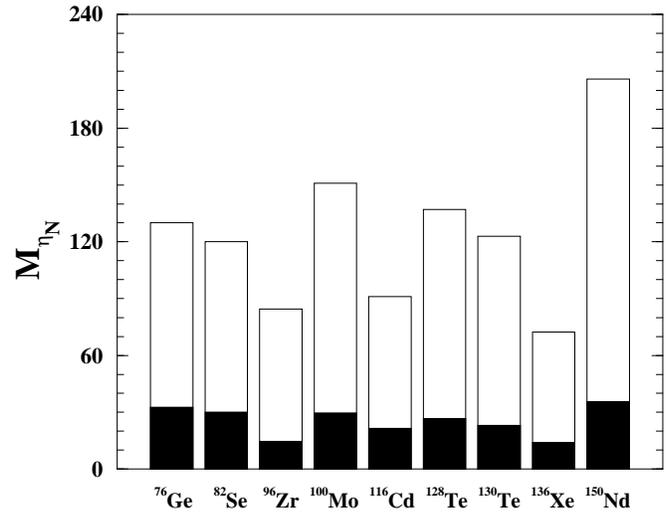
a given nucleus $Y$ to the light neutrino mass, heavy neutrino mass
and Majoron signals, respectively.}
\label{table.2}
\begin{tabular}{lrrrrrrrrr}
\multicolumn{10}{c}{ $(\beta\beta)_{0\nu}-decay: 
0^{+}\rightarrow{0^{+}}$ 
 transition} \\ \cline{2-10}
M. E. & $^{76}Ge$ & $^{82}Se$ & $^{96}Zr$ & $^{100}Mo$ &
 $^{116}Cd$ & $^{128}Te$ & $^{130}Te$ & $^{136}Xe$ & $^{150}Nd$ 
\\ \hline
\multicolumn{10}{c}{ light Majorana neutrino  (I=light)} \\
$M_{VV}^I$ &
 0.80  &   0.74  &   0.45  &   0.82  &  0.50  &
 0.75  &   0.66  &   0.32  &   1.14 \\
$M_{AA}^I$ &
 2.80  &   2.66  &   1.54  &   3.30  &  2.08  &
 2.21  &   1.84  &   0.70  &   3.37 \\
$M_{PP}^I$ &
 0.23  &   0.22  &   0.15  &   0.26  &  0.15  &
 0.24  &   0.21  &   0.11  &   0.35 \\
$M_{AP}^I$ &
-1.04  &  -0.98  &  -0.65  &  -1.17  &  -0.69 &
-1.04  &  -0.91  &  -0.48  &  -1.53 \\
 & & & & & & & & & \\
$M_{VV}^I + M_{AA}^I$ &
 3.60  &   3.40  &   1.99  &   4.12  &   2.58 &
 2.96  &   2.50  &   1.02  &   4.51 \\
$M_{<m_\nu >}^I$ &
 2.80  &   2.64  &   1.49  &   3.21  &   2.05 &
 2.17  &   1.80  &   0.66  &   3.33 \\
 & & & & & & & & & \\
\multicolumn{10}{c}{ heavy Majorana neutrino  (I= heavy)} \\
$M_{VV}^I$ &
 23.9  &   22.0  &   16.1  &   28.3  &   17.2 &
 25.8  &   23.4  &   13.9  &   39.4 \\
$M_{MM}^I$ &
-55.4  &  -51.6  &  -38.1  &  -67.3  &  -39.8 &
-60.4  &  -54.5  &  -31.3  &  -92.0 \\
$M_{AA}^I$ &
 106.  &   98.3  &   68.4  &   123.  &   74.0 &
 111.  &   100.  &   58.3  &   167. \\
$M_{PP}^I$ &
 13.0  &   12.0  &   9.3   &   16.1  &   9.1  &
 14.9  &   13.6  &   7.9   &   23.0 \\
$M_{AP}^I$ &
-55.1  &  -50.7  &  -41.1  &  -70.1  &  -39.0 &
-64.9  &  -59.4  &  -34.8  &  -101. \\
 & & & & & & & & & \\
$M_{VV}^I + M_{AA}^I$ &
 130.  &   120.  &   84.5  &    151. &   91.1 &
 137.  &   123.  &   72.3  &  206. \\
$M_{\eta_{_N}}^I$ &
 32.6  &  30.0   &   14.7  &   29.7  &   21.5 &
 26.6  &  23.1   &   14.1  &   35.6 \\
 & & & & & & & & & \\
\multicolumn{10}{c}{sensitivity to neutrino mass signal} \\
$G_{01} \times 10^{15} y$ & 
7.93 & 35.2 & 73.6 & 57.3 & 62.3 & 2.21 & 55.4 & 59.1 & 269. \\ 
$\zeta_{<m_\nu >} (Y)$ &
 2.49 &  4.95 & 4.04 &  7.69 &  5.11 &
 1.02 &  4.24 & 1.60 &  17.3 \\
$\zeta_{\eta_{_N}} (Y)$  &
 2.90 &  5.64 & 3.98 &  7.10 &  5.36 &
 1.25 &  5.45 & 3.43 &  18.5 \\
 & & & & & & & & & \\
\multicolumn{10}{c}{sensitivity to Majoron signal} \\
$G_{B} \times 10^{17} y$ & 
7.40 & 62.3 & 159. & 106. & 104. & 0.59 & 79.6 & 82.8 & 640. \\
$\zeta_{<g>} (Y)$  &
 2.41 &  6.59 & 5.93 & 10.5 &  6.60 &
 0.53 &  5.08 & 1.90 &  26.7 \\
\end{tabular}
\end{table}

\begin{table}[t]
\caption{The present state of the
Majorana neutrino mass and Majoron searches in
$\beta\beta$-decay experiments.
$T^{exp-0\nu}_{1/2}$(present) and $T^{exp-0\nu \phi}_{1/2}$(present) 
are the best presently available lower limit on
the half-life of the $0\nu\beta\beta$-decay  and $0\nu\beta\beta\phi$-decay 
for a given isotope, respectively.
The corresponding upper limits on lepton number non-conserving parameters 
${<m_\nu >}$, ${<g>}$  and $\eta_{N}$  are presented. 
$T^{exp-0\nu}_{1/2}$($<m_\nu >^{best}$), 
$T^{exp-0\nu}_{1/2}$($\eta^{best}_{_N}$) and 
$T^{exp-0\nu\phi}_{1/2}$($<g>^{best}$) 
are calculated half-lifes of $0\nu\beta\beta$-decay
assuming  $<m_\nu > = <m_\nu >^{best}$,  
$\eta_{_N} = \eta_{_N}^{best}$ and $<g> = <g>^{best}$, respectively.
Here,
$<m_\nu >^{best} = 0.62$ eV $\eta_{_N}^{best} = 1.0\times 10^{-7}$ 
and $<g >^{best} = 6.9\times 10^{-5}$ are the best limits deduced from 
the  $^{76}Ge$ [38] and $^{128}Te$  [43] experiments.
}
\label{table.3}
\begin{tabular}{llllll}
Nucleus  & $^{76}Ge$ & $^{82}Se$ & $^{96}Zr$ & $^{100}Mo$ & $^{116}Cd$ \\
\hline
 $T^{exp-0\nu}_{1/2}$(present) [y] &
 $ 1.1\times 10^{25}$    & $  2.7\times 10^{22}$ &
 $ 3.9\times 10^{19}$    & $  5.2\times 10^{22}$ &  $ 2.9\times 10^{22}$ \\
Ref. & \cite{hdmo97} & \cite{ell92} & \cite{kaw93} & \cite{eji96} &
       \cite{dane95} \\ 
 $<m_\nu >$ [eV]  & 0.62 & 6.3 & 203. & 2.9 & 5.9 \\
 $T^{exp-0\nu}_{1/2} (<m_\nu >^{best})$ [y] &  
 $1.1\times 10^{25}$ & $2.8\times 10^{24}$ & $4.2\times 10^{24}$ &
 $1.2\times 10^{24}$ & $2.6\times 10^{24}$ \\
 $\eta_{_N} $ & $1.0\times 10^{-7}$ & $1.1\times 10^{-6}$ & 
 $4.0\times 10^{-5}$ &
 $6.2\times 10^{-7}$ & $1.1\times 10^{-6}$ \\   
 $T^{exp-0\nu}_{1/2} (\eta_{_N}^{best})$ [y] &  
 $1.1\times 10^{25}$ & $2.9\times 10^{24}$ & $5.8\times 10^{24}$ &
 $1.8\times 10^{24}$ & $3.2\times 10^{24}$ \\
 & & & & & \\ 
 $T^{exp-0\nu\phi}_{1/2}$(present) [y] &
 $ 7.9\times 10^{21}$    & $  1.6\times 10^{21}$ &
 $ 3.9\times 10^{19}$    & $  5.4\times 10^{21}$ &  $ 1.2\times 10^{21}$ \\
Ref. & \cite{gue97} & \cite{moe94} & \cite{kaw93} & \cite{eji96} &
       \cite{dane95} \\ 
 $<g> $ & $4.7\times 10^{-4}$ & $3.8\times 10^{-4}$ & $2.7\times 10^{-3}$ &
 $1.3\times 10^{-4}$ & $4.4\times 10^{-4}$ \\   
 $T^{exp-0\nu\phi}_{1/2} (\eta_N^{best})$ [y] &  
 $3.7\times 10^{23}$ & $4.9\times 10^{22}$ & $6.0\times 10^{22}$ &
 $1.9\times 10^{22}$ & $4.9\times 10^{22}$ \\
 & & & & & \\
\hline
Nucleus & $^{128}Te$ & $^{130}Te$ & $^{136}Xe$ & $^{150}Nd$ & \\
\hline
 $T^{exp-0\nu}_{1/2}$(present) [y] &
 $  7.7\times 10^{24}$   &  $  8.2\times 10^{21}$  &
 $  4.2\times 10^{23}$   &  $  1.2\times 10^{21}$  &  \\
Ref.  & \cite{bern92} & \cite{ale94} & \cite{bus96} & \cite{sil97} & \\
 $<m_\nu >$ [eV]  & 1.8 & 13. & 4.9 & 8.5  \\
 $T^{exp-0\nu}_{1/2} (<m_\nu >^{best})$ [y] &  
 $6.6\times 10^{25}$ & $3.8\times 10^{24}$ & $2.7\times 10^{25}$ &
 $2.3\times 10^{23}$ \\
 $\eta_N $ & $2.9\times 10^{-7}$ & $2.0\times 10^{-6}$ & $4.5\times 10^{-7}$ &
 $1.6\times 10^{-6}$ \\   
 $T^{exp-0\nu}_{1/2} (\eta_{_N}^{best})$ [y] &  
 $5.9\times 10^{25}$ & $3.1\times 10^{24}$ & $7.9\times 10^{24}$ &
 $2.7\times 10^{23}$ \\
 & & & & & \\ 
 $T^{exp-0\nu\phi}_{1/2}$(present) [y] &
 $  7.7\times 10^{24}$   &  $  2.7\times 10^{21}$  &
 $  1.4\times 10^{22}$   &  $  2.8\times 10^{20}$  &  \\
Ref.  & \cite{bern92} & \cite{bern92} & \cite{bus96} & \cite{sil97} & \\
 $<g> $ & $6.9\times 10^{-5}$ & $3.8\times 10^{-4}$ & $4.5\times 10^{-4}$ &
 $2.2\times 10^{-4}$ \\   
 $T^{exp-0\nu\phi}_{1/2} (\eta_{_N}^{best})$ [y] &  
 $7.7\times 10^{24}$ & $8.2\times 10^{22}$ & $5.9\times 10^{23}$ &
 $3.0\times 10^{21}$ \\
\end{tabular}
\end{table}


\begin{figure}
\caption{Calculated light and heavy neutrino exchange 
$0\nu\beta\beta$-decay nuclear matrix elements for the A=76 and 128 systems.
The partial matrix elements $M_{VV}$, $M_{AA}$, $M_{MM}$, $M_{PP}$ and
$M_{AP}$ originate from vector, axial-vector, weak magnetism, pseudoscalar
coupling and the interference of the axial-vector and induced 
pseudoscalar coupling,
 respectively. $M_{<m_\nu >}$ and $M_{\eta_{_N}}$ are 
$0\nu\beta\beta$-decay matrix elements associated with the $<m_\nu >$ and
$\eta_{_N}$ parameters, respectively.  
}
\label{fig.1}
\end{figure}

\begin{figure}
\caption{Calculated nuclear matrix elements $M_{<m_\nu >}$, $M_{\eta_{_N}}$ 
[see Eqs. (3) and (13)-(18)], 
sensitivities $\zeta_{<m_\nu >}$, $\zeta_{\eta_{_N}}$ and
$\zeta_{<g>}$ 
for the experimentally interesting A=76, 82, 96, 100, 116, 128,
130, 136 and 150 nuclear systems. 
$\zeta_{<m_\nu >}$, $\zeta_{\eta_{_N}}$ and
$\zeta_{<g>}$ are 
sensitivities to light neutrino mass, heavy neutrino
mass and Majoron signal [see Eq. (34) ], respectively.
The open and black bars  correspond to
results obtained without and with the inclusion of the  
pseudoscalar interaction and of weak-magnetism.}
\label{fig.2}
\end{figure}

\begin{figure}
\caption{The sensitivity of different experiments to the lepton-number
violating parameters $<m_\nu >$, $\eta_{_N}$ and $<g>$ 
are illustrated by histograms on the left side. 
The best presently available lower limits on the $0\nu\beta\beta$-decay
half-life $T^{exp-0\nu}_{1/2}$ and $0\nu\beta\beta\phi$-decay 
$T^{exp-0\nu \phi}_{1/2}$ are displayed by black bars
in histograms on the right side. The open bars in these histograms 
indicate the half-life limits $T^{exp-0\nu}_{1/2}$($<m_\nu >^{best}$), 
$T^{exp-0\nu}_{1/2}$($\eta^{best}_{_N}$) and 
$T^{exp-0\nu\phi}_{1/2}$($<g>^{best}$) to be reached by a given experiment
to reach the presently best limit on $<m_\nu >$, $\eta_{_N}$ and $<g>$,
respectively.}
\label{fig.3}
\end{figure}

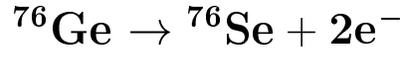
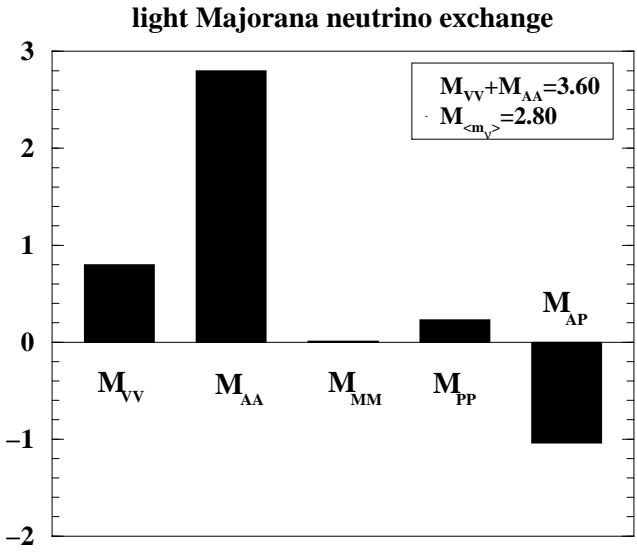
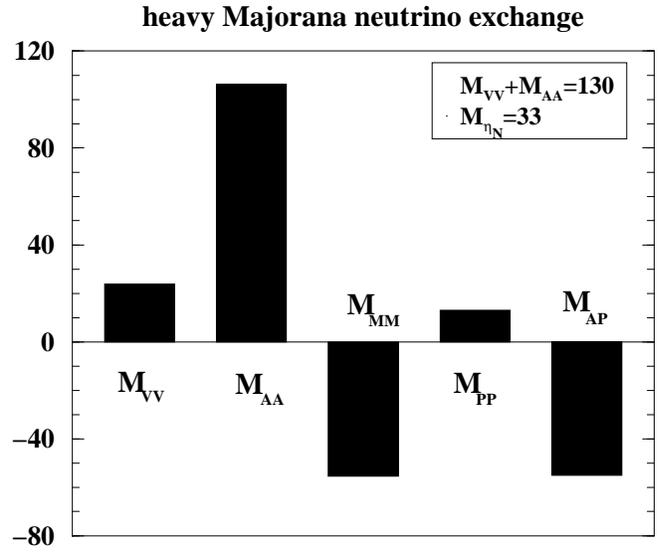
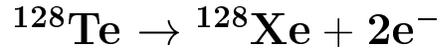
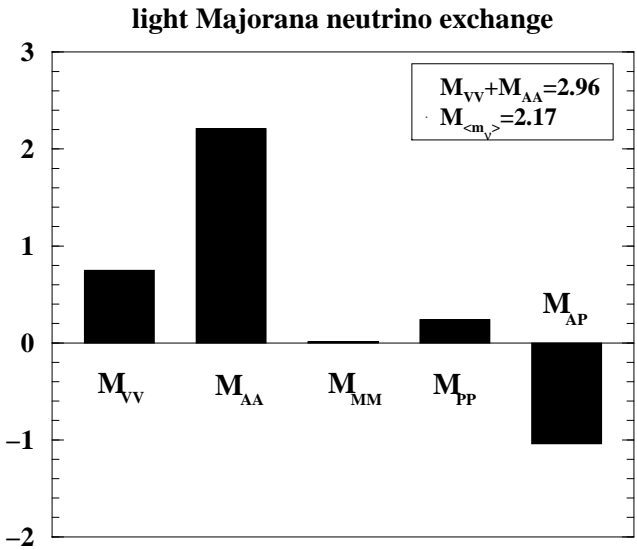
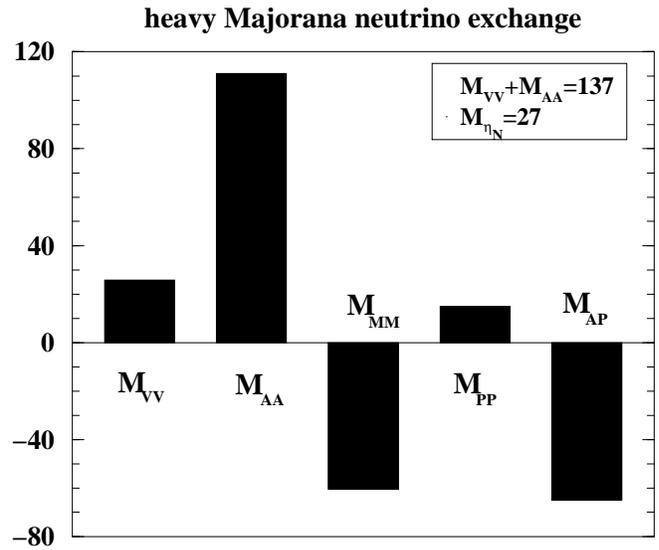
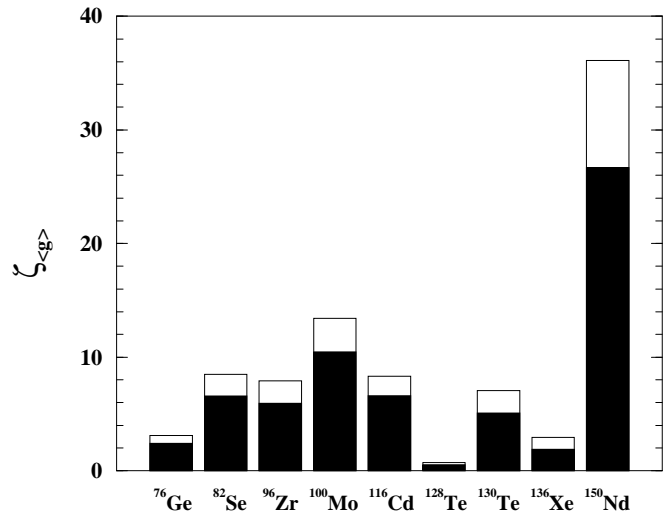
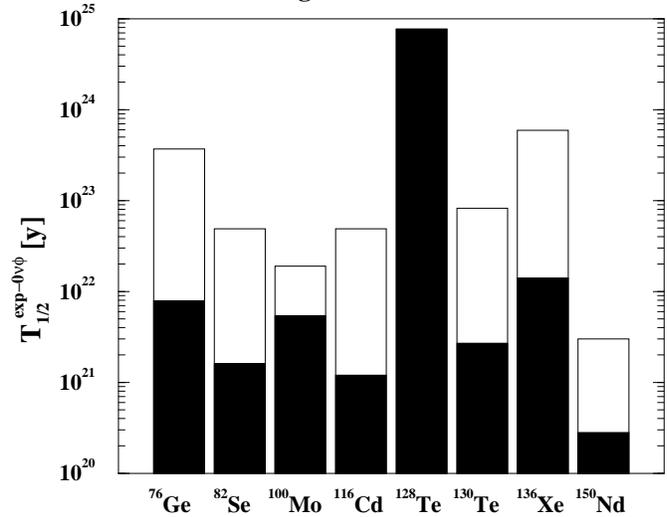
\end{document}